\documentclass{elsart}
\usepackage{amsmath, amssymb, epsfig}
\begin{document}

\begin{frontmatter}

\title{rf-studies of vortex dynamics in isotropic type-II superconductors.}

\author{N. L\"utke-Entrup, B. Pla\c{c}ais, P. Mathieu and Y. Simon},

\address{ Laboratoire de Physique de la Mati\`ere Condens\'ee de
l'Ecole Normale Sup\'erieure,\\
associ\'e au  CNRS et aux Universit\'es Paris 6 et 7,
24 rue Lhomond, F-75231 Paris Cedex 05}

\begin{abstract}
We have measured the surface impedance of thick superconductors in the
mixed state over a broad 2\thinspace kHz -- 20\thinspace MHz frequency
range.  The depinning cross-over is observed; but it is much broader
than expected from classical theories of pinning.  A striking result
is the existence of size effects which invalidate the common
interpretation of the low-frequency surface inductance in terms of a
single penetration depth.  Instead, a two-mode description of vortex
dynamics, assuming free vortex flow in the bulk and surface pinning,
accounts quantitatively for the spectrum of the complex apparent
penetration depth.
\end{abstract}


\end{frontmatter}

\section{Introduction}\label{Introduction}

The presence of an array of quantized vortices ($\psi\!=\!0$ lines) in
a type-II superconductor entails a frictional force in the Euler
equation for the supercurrent.  For an isotropic material it takes the
form \cite{Mathieu88}
\begin{equation}
\boldsymbol E + 
\frac{m}{e}\frac{\partial\boldsymbol V_{\!\!{\rm s}}}{\partial t} = 
\beta\boldsymbol\omega\times\left[(\boldsymbol J_{\!{\rm s}}+\text{curl}\,
\boldsymbol\varepsilon)\times\boldsymbol\nu\right] + \beta'
\boldsymbol\omega\times(\boldsymbol J_{\!{\rm s}} +
\text{curl}\,\boldsymbol\varepsilon) =
\boldsymbol\omega\times\boldsymbol v_{\rm L},
\label{LondonAC}
\end{equation}
where $\boldsymbol V_{\!\!{\rm s}}$, $\boldsymbol J_{\!{\rm
s}}=-n_{\rm s}e\boldsymbol V_{\!\!{\rm s}}$ and $n_{\rm s}$ are
coarse-grained mean values of the superfluid velocity, supercurrent,
and density of superelectrons; $m$ and $e$ are the electronic mass and
charge.  The electric field $\boldsymbol E$ is directly related to the
vortex field $\boldsymbol\omega=n_{\rm V}\varphi_0\boldsymbol\nu$ and
the line velocity $\boldsymbol v_{\rm L}$ through the generalized
Josephson equation.  The vector $\boldsymbol\omega$ incorporates both
density $n_{\rm V}$ and orientation $\boldsymbol\nu$ of the vortex
lines.

The macroscopic London equation,
\cite{Abrikosov65,Hocquet92}
\begin{equation}
\boldsymbol\omega = \boldsymbol B -
\frac{m}{e}\text{curl}\,\boldsymbol V_{\!\!{\rm s}} = \boldsymbol B +
\text{curl}\,(\mu_0\lambda^2\boldsymbol J_{\!{\rm s}}),
\label{LondonDC}
\end{equation}
shows that vortex lines may deviate from magnetic field lines
($\boldsymbol\omega\neq\boldsymbol B$) in the presence of a
supercurrent.  In Eq.~(\ref{LondonDC}) $\lambda$ is a field-dependent
penetration depth.

Here we consider a slab in a perpendicular applied field $\boldsymbol
B_0$. The vortex density $n_{\rm V}\simeq B_0/\varphi_0$ is nearly
constant, but $\boldsymbol\nu$ may be quite different from
$\boldsymbol B/B$; such bending effects are particularly important
near the sample surface.

Eq.~(\ref{LondonAC}) is the superconducting analog of the acceleration
equation for rotating superfluids (Eq.(24) in
Ref.~\cite{Bekarevich61}) and has been written in its most general
form such as deduced from conservation laws.  The line tension
$\boldsymbol\varepsilon=\varepsilon\boldsymbol\nu$ is the
thermodynamic conjugate variable of $\boldsymbol\omega$.  Kinetic
coefficients $\beta $ and $\beta'$ describing the longitudinal
(resistive) and transverse (reactive) components of the friction force
are analogs of the mutual friction parameters $B$ and $B'$ in rotating
superfluids \cite{Donnelly91,Hook97}.  Allowing for a normal-current
contribution $\boldsymbol J_{\!{\rm n}}=\bar\sigma\boldsymbol E$
($\sigma=\bar\sigma_\parallel, \sigma^\prime=\bar\sigma_\perp$) in the
total current $\boldsymbol J \!=\!\boldsymbol 
J_{\!{\rm n}}\!+\!\boldsymbol J_{\!{\rm s}}$, $\beta $ is related to the
longitudinal flux-flow conductivity $\sigma_{\rm f}\!=\!
\sigma\!+\!1/\beta\omega$. The Hall-angle is related to $\beta'$
through $\tan\theta_H\!=\!(\sigma'\!+\!1/\beta'\omega )/\sigma_{\rm
f}$.

In conventional dirty superconductors dissipation ($\beta$) arises
from normal-current back-flow and relaxation phenomena at the vortex
core, while Hall angles are small.  Of special interest for future
experiments will be the case of clean superconductors at low
temperatures where $\sigma_{\rm f}\simeq 1/\beta\omega$ and $\theta_H$
are expected to be governed by the discreteness of core levels in
relation with the symmetry of the order parameter
\cite{Caroli64,Kopnin97}.

Vortex dynamics is best probed by measuring the surface impedance
{$Z\!=\!R-{\rm i} X$} in the linear regime, which is conveniently
expressed in terms of an effective complex penetration depth
\begin{equation}
\lambda_{\rm ac}=-\frac{Z}{{\rm 
i}\mu_0\Omega}=\lambda^\prime+{\rm i}\lambda^{\prime\prime}.
\label{Zorro}\end{equation}
A perfect sample, completely free from defects, should behave like a
continuous resistive medium with $Z=(1+{\rm i})\delta_{\rm f}/2$,
where $\delta_{\rm f}=(2/\mu_0\sigma_{\rm f}\Omega)^{1/2}$ is the
usual skin-depth.  In real samples, this ideal response has been
observed in conventional superconductors at microwave frequencies
$\Omega/2\pi\sim 10$\thinspace GHz \cite{Rosenblum64}, and also for
moving vortex arrays driven by a large superimposed dc transport
current ($I\gg I_{\rm C}$) \cite{Vasseur97}.  At low frequencies (and
$I=0$) the linear ac response is strongly altered by pinning, so that
a sample in the mixed state behaves as a true loss-free
superconductor, with $\lambda^{\prime\prime}\simeq0$
\cite{Gittleman66} and a quasistatic penetration depth
$\lambda^\prime\sim$ 1--100\thinspace $\mu$m \cite{Campbell69}. The
way pinning and flux flow interfere in the cross-over regime
($\Omega/2\pi\sim$ 0.1--100\thinspace MHz in dirty conventional
materials) is involved in the complex spectrum $Z(\Omega)$, which
consequently should be the main concern for the dynamics of pinning
(see the theoretical review \cite{Beek93} and references therein).
From a practical point of view it is equally important to know the
width of the cross-over regime and the position of its center point,
the depinning-frequency $\Omega_{\rm p}$, because they determine both
the onset of losses at low frequencies and the recovery of the free
flux-flow response at high frequencies.

Our experiment is designed to study the skin-depth spectrum
$\lambda_{\rm ac}(\Omega)$ in the cross-over regime. For this purpose,
we measure the penetration of small rf magnetic fields into thick
superconducting slabs in a large frequency range, $\Omega/2\pi\sim$
2\thinspace kHz--20\thinspace MHz .

The quasistatic regime is classically described in terms of an elastic
interaction between the vortex array and sample defects
\cite{Tinkham96} (p.370). Assuming the existence of a pinning force
density, small vortex displacements $\boldsymbol u $ about their
equilibrium position result from a Lorentz force $\boldsymbol J\times
\boldsymbol\omega$ per unit volume, balanced by a linear restoring
force $\boldsymbol F_{\!{\rm p}}\!=\! - \alpha_{\rm L}\boldsymbol u$,
where $\alpha_{\rm L}$ is the phenomenological Labusch parameter.
With increasing frequency a viscous term $-\eta\boldsymbol v_{\rm L}$
($\eta =\sigma_{\rm f}\omega^2$) comes into play and damps vortex
motion.  Linearizing both the Josephson relation $\boldsymbol E =
\boldsymbol\omega\times \boldsymbol v_{\rm L} $ ($\boldsymbol v_{\rm
L}=\dot{\boldsymbol u}$) and the force-balance equation, one obtains a
local constitutive equation in the form of the Ohm's law $\boldsymbol
J_{\!\perp}=\sigma^*\boldsymbol E$, where $\sigma^*=\sigma_{\rm
f}\!+\!{\rm i}\alpha_{\rm L}/(\omega^2\Omega)$.  This results in a
one-mode electrodynamics with a complex penetration depth
$\lambda_{\rm ac}$ given by
\begin{equation} 
\frac{1}{\lambda_{\rm ac}^{2}} = -{\rm i}\mu_0\sigma^*\Omega =
\frac{1}{\lambda_{\rm C}^2} - \frac{2{\rm i}}{\delta_{\rm f}^2}.
\label{Campbell}\end{equation}
Here $\lambda_{\rm C}=\omega/(\mu_0\alpha_{\rm L})^{1/2}$ is the
Campbell penetration depth for small quasistatic vortex motions
\cite{Campbell69}.  The spectrum (\ref{Campbell}) has been amended so
as to include thermally-assisted vortex motion (flux creep)
\cite{Coffey91,Koshelev91}.  Thermal activation entails an additional
$\Omega$-dependence in $\alpha_{\rm L}$ (or $\lambda_{\rm C}$) at low
frequencies, but it preserves the electrodynamic description in terms
of a \emph{local effective conductivity} $\sigma^*$ \cite{Beek93}.
  
\begin{figure}[ht]
\begin{center}
\includegraphics[width=0.7\linewidth]{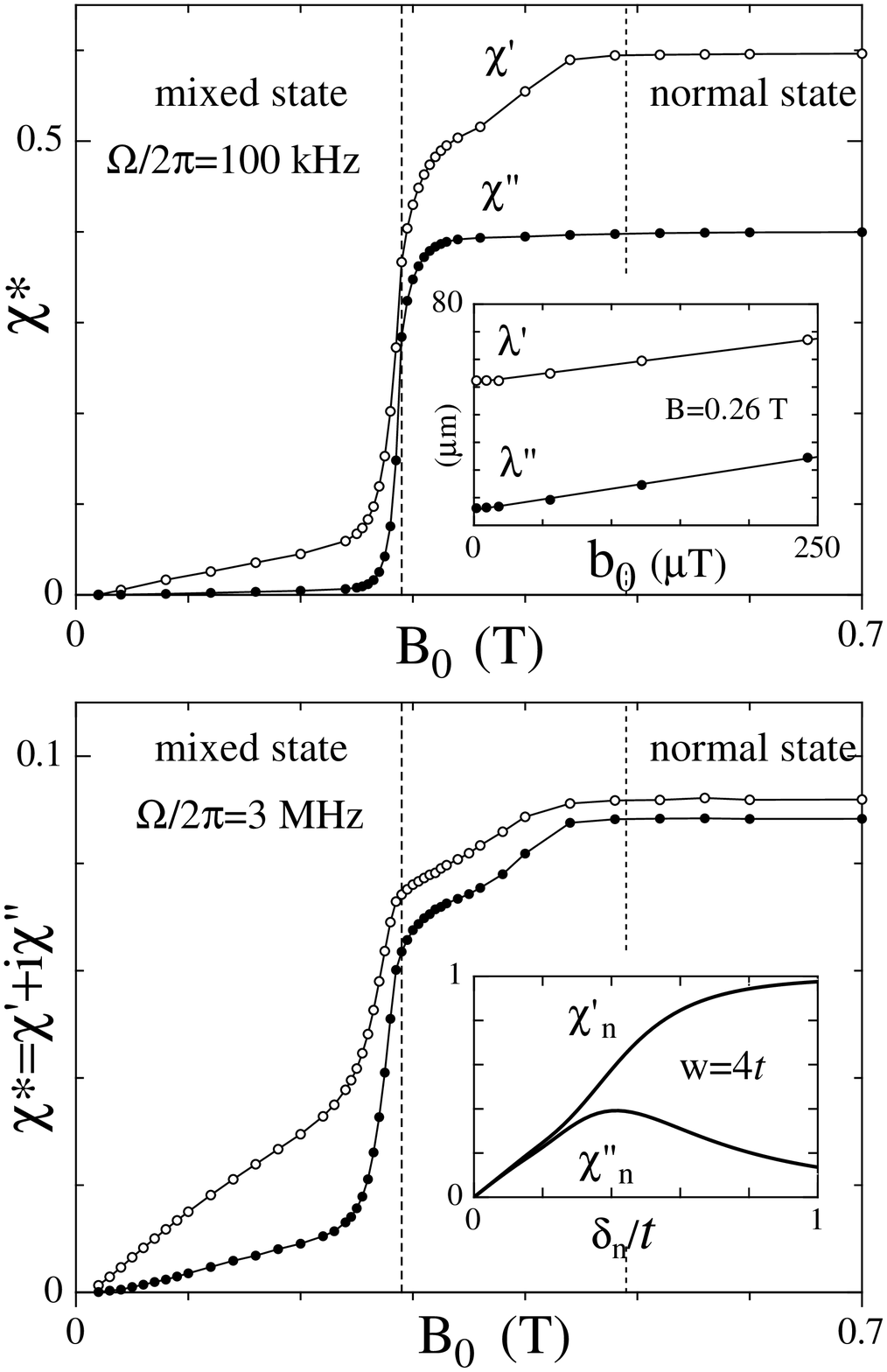}
\bigskip
\caption{Transverse susceptibility of a $6.1(w)\times
1.22(t)$\thinspace mm$^2$
PbIn${10.5\%wt}.$ slab ($\rho_{\rm n}=10.3\,\mu\Omega$cm) in a broad
range of normal fields including the mixed state ($B_0\leq0.29\,$T) and
the surface sheath regime ($0.29\,$T$\leq B_0\leq 0.49\,$T). The upper
panel shows the quasistatic response at low frequency with vanishing
losses ($\chi^{\prime\prime}$) and a finite in-phase apparent
penetration depth $\lambda^\prime=\chi^\prime t/2$.  In the lower
panel, $\Omega$ is close to the depinning threshold $\Omega_{\rm p}$,
and $\chi^{\prime\prime}$ is clearly visible.  For each frequency, the
normal state response (plateau for $B_0\geq0.49\,$T) is used to
calibrate the phase and the amplitude of the excitation by comparison
with a two-dimensional calculation (inset of the lower panel and
Eqs.~(\ref{2DnormalChi})).  Non-linear effects are displayed
in the inset of the upper figure: strictly the linear response
requires extrapolating data to $b_0=0$ for each field and frequency;
in practice we work with a constant but small level $b_0\sim 1\,\mu$T
(or induced currents $K=b_0/\mu_0\sim 10\,$mA/cm), which translates
into minute vortex displacements $u\simeq\lambda^\prime 
b_0/B_0\sim 1$\,\AA. }
\label{fieldtransition}\end{center}\end{figure}

According to a recent \emph{two-mode electrodynamics}
\cite{Sonin92,Placais96}, which follows from linearized
Eqs.~(\ref{LondonAC}) and (\ref{LondonDC}), there is, apart from the
ordinary flux-flow mode (FF-mode), a second mode associated with
strain and corresponding to $\text{curl}\,\boldsymbol\varepsilon$
terms in Eq.~(\ref{LondonAC}): this vortex-strain mode (VS-mode) is
evanescent and non dissipative.  It is worth noting that in
Eq.~(\ref{LondonAC}) (like for superfluids in \cite{Bekarevich61})
shear effects are disregarded.  We adopt a rather old line of
reasoning \cite{DeGennes64}, according to which the small shear
rigidity of the vortex array may be ignored in most dynamic problems.
It remains that the shear modulus $c_{66}$ is essential for questions
about the detailed equilibrium configuration of the vortex lattice; we
return to this point in Sec.~\ref{conclusion}.  On the other hand one
cannot ignore the role of the VS-mode: although it dies off over a
short depth $\lambda_{\rm V}\gtrsim a$ from the surface, it can carry
large non-dissipative currents, $\boldsymbol J_{\!{\rm
s}}=-\text{curl}\,\boldsymbol\varepsilon$ ($J_{\rm s}\lesssim
\varepsilon/\lambda_{\rm V}\sim 10^7$ A/cm$^2$).  On introducing the
appropriate surface condition for small reversible vortex motion, the
amplitude and the phase of the two modes, and ultimately the surface
impedance of the sample, are determined. It turns out that for the ac
response of a perfect sample the surface condition
$\boldsymbol\nu_\parallel=0$ applies, reducing the amplitude of the
VS-mode to its (small) diamagnetic contribution \cite{Vasseur97}.
However, as suggested recently \cite{Sonin92,Entrup97}, the VS-mode
can be considerably enhanced by surface roughness: surface
irregularities on the scale of $a$ allow for an hysteresis in the
``contact angle'' between the vortex array and the averaged smoothed
sample surface ($\nu_\parallel\lesssim 0.1$).  On substituting the
vortex slippage condition
$\boldsymbol\nu_\parallel=-\boldsymbol{u}/l$, for the ideal boundary
condition, we find a number of important results. The hallmark of the
two-mode electrodynamics is the original shape of the $\lambda_{\rm
ac}(\Omega)$ spectrum (see Eq.~(\ref{admittance}) and Figs.~\ref
{depinningspectrum} and \ref{PotPourritArgand} below).  Furthermore,
we predict unexpected size-effects which are discussed in
Sec.~\ref{size_effects} and Fig.~\ref{thinslablossess}.
 
In this paper we detail the experimental procedure
(Sec.~\ref{set-up}), show new results on PbIn alloys over an extended
frequency range both for perpendicular and oblique fields, and discuss
finite-size effects at some length. A preliminary investigation of the
sheath-superconductivity regime (above $B_{\rm C2}$), included in
Sec.~\ref{sheath_superconductivity}, supports our interpretation of the
mixed-state spectra in term of two modes. The paper ends with a
discussion about pinning mechanisms.
 
\section{Experimental principle}\label{set-up}
 
We have investigated a series of long superconducting slabs (length
$L$ in the $x$-direction, width $w$ in the $y$-direction).  The
applied magnetic field $B_0$, in the $yz$-plane, makes an angle
$\theta$ with the $z$-direction. We concentrate to the $\theta=0$ and
$\theta=\pi/4$ orientations referred to below as normal and oblique
field.  A small ac ripple $b_0e^{-i\Omega t}$ along the $x$-axis is
produced by a 10--100 turns excitation coil. Induced currents make the
vortices oscillate in the $xz$ planes. The ac flux $\Phi^*e^{-i\Omega
t}$ through the cross-section $w\times t$ is detected by means of a
10--100 turns pick-up coil wound on the sample and measured with
phase-sensitive lock-in analysers (EGG 5202 and 5302) in the broad
$\Omega/2\pi=$2\thinspace kHz--20\thinspace MHz frequency range.  In
order to obtain a high resolution in the loss-angle
$\varphi\!=\!\arctan(\lambda^{\prime\prime}/\lambda^\prime)$
($\delta\varphi\lesssim0.3$ degree below 1\thinspace MHz) we used thin
bronze wire to make pick-up coils transparent to the ac-field, and the
rf chamber was enclosed in a high-conductivity copper can.  At the low
excitation levels of the linear regime ($b_0\lesssim 1\,\mu$T in
Fig.~\ref{fieldtransition}), the resolution was ultimately limited to
$\delta\lambda\lesssim 0.1\,\mu$m by electronic noise.  Flux-flow
conductivities and critical currents were measured beforehand from dc
voltage-current characteristics ($I\leq 25$ A).

Most measurements reported in this paper were carried out on
PbIn${10.5\% wt}$ slabs $40 (L)\times 6 (w)\times t$ mm$^3$ with
$t=0.3, 1.3, 4$ mm.  Homogeneous lead-indium alloys are easily
prepared from pure metals, and extensive measurements in the
literature give reliable values of their $B_{\rm C2}$, $\kappa$ and
$\rho_{\rm n}$ ($10.3\,\mu\Omega$cm) \cite{Farrel69}.  Slabs were
spark-cut directly from ingots and optionally chemically etched.  With
$\kappa=3.5$ magnetization effects are negligible down to very low
fields, and $\omega=B_0$.  On the other hand, $B_{\rm C2}$
(0.29\thinspace T at 4.2\thinspace K) is low enough so that the
field-phase diagram can be investigated up to the normal state with
our 0.8\thinspace Tesla magnet (see Fig.~\ref{fieldtransition}).
Measurements on very clean niobium ($\rho_{\rm n}\lesssim 10$
n$\Omega$cm), first reported in Ref.~\cite{Entrup97}, will be detailed
elsewhere.  We show in Fig.~\ref {PotPourritArgand} new data on
PbIn${5\% wt}$ ($\rho_{\rm n}=6.5\,\mu\Omega$cm) and standard vanadium
($99.5\%$ with $\rho_{\rm n}=0.6\,\mu\Omega$cm), which exhibit the same
behaviour as PbIn${10.5\% wt}$.

For the sake of obtaining absolute values of $\lambda_{\rm ac}$ we
need to calibrate the ac-field $b_0$ accurately both in amplitude and
phase.  Several techniques can be used for this purpose: the phase can
be obtained from that of the current in the rf excitation coil, or by
measuring the magnetic flux either through an auxiliary pick-up coil,
or through the gap between the pick-up coil and the sample in the
Meissner state $B_0=0$.  The Meissner state is used throughout this
work as the reference $\Phi^*=0$ state, neglecting the penetration
over the small London depth $\lambda_{\rm L}\sim0.1\,\mu$m.  The most
accurate technique, however, makes use of the complex susceptibility of
the normal state $\chi^*_{\rm n}$.  Knowing $\rho_{\rm n}$, the flux
distribution in an isotropic normal metal can be easily calculated
thanks to the homogeneous boundary conditions achieved in this
geometry.  Note in this respect that the situation is far less
tractable in resistivity experiments where the spatial distribution of
the ac applied current is an intricate problem; moreover, the measured
voltage strongly depends on the leads arrangement.

For a precise calibration of the phase of the exciting field in the
normal state, we need the exact two-dimensional solution for the field
distribution in a rectangular cross section $w\times t$, which reads :
\begin{gather*} 
\frac{b}{b_0} = \sum_{m,n\;{\rm odd}}\frac{16}{m n \pi^2}
\frac{k_{mn}^2}{k_{mn}^2-2i/\delta_{\rm n}^2} \sin\frac{m\pi y}{w}
\sin\frac{l\pi z}{t}\\ (k_{mn}^2 =
\frac{m^2\pi^2}{w^2}+\frac{n^2\pi^2}{t^2}).
\end{gather*}
Integration over the section yields the apparent susceptibility
$\chi_{\rm n}^*=\Phi_{\rm n}^*/b_0wt$ :
\begin{equation} 
\chi_{\rm n}^* = \sum_{m,n\;{\rm odd}}\;\frac{64}{m^2 n^2
\pi^4}\;\frac{k_{mn}^2}{k_{mn}^2-2i/\delta_{\rm n}^2}.
\label{2DnormalChi}\end{equation}
The inset in the lower Fig.~\ref{fieldtransition} shows $\chi_{\rm
n}^\prime$ and $\chi_{\rm n}^{\prime\prime}$ as a function of
$\delta_{\rm n}/t\propto1/\sqrt{\Omega}$, calculated from
Eq.~(\ref{2DnormalChi}) for $w=4t$.  As a cross check we always
observed that the phase determined by other methods coincides with
that obtained from the normal-state response within 0.3 degrees below
1\thinspace MHz. The calibration from the normal-state response is,
for a large part, unaffected by stray capacitances from which coil
circuits usually suffer at high frequencies.

Whereas a 2D calculation was necessary in the normal-state response,
the situation is different in the mixed state because of the strong
anisotropy of the critical currents and the flux flow resistivity.  In
particular, as can be checked experimentally by making $\theta=\pi/2$,
the flux penetration through the lateral faces parallel to the
vortices is negligible below $0.9 B_{\rm C2}$.  Flux penetration in
normal field can thus be regarded as a one-dimensional problem with
$\Phi^*=2w b_0\lambda_{\rm ac}$.  Anyway, the question of the lateral
penetration can be circumvented by working in oblique field where the
4 faces play symmetric roles so that $\Phi^*=2(w+t) b_0\lambda_{\rm
ac}$.  In oblique field, bulk currents flow at an angle $\theta=\pi/4$
with vortices, and the effective resistivity is $\rho_{\rm
f}(\theta)=\rho_{{\rm f}\perp} \cos^2\theta$.

\begin{figure}[ht]
\begin{center}\leavevmode
\includegraphics[width=1\linewidth]{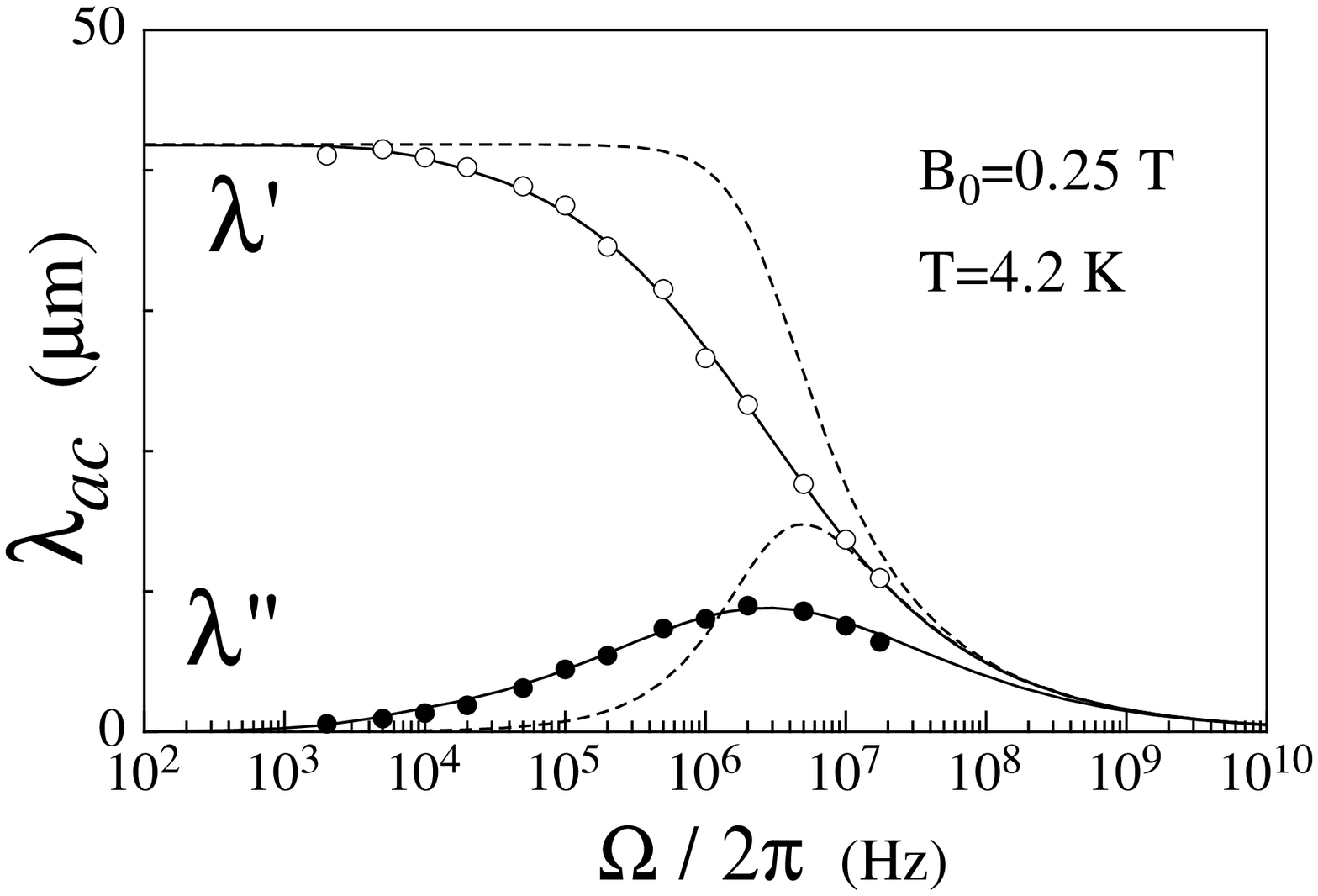}
\bigskip
\caption{Broad-band spectrum of the apparent penetration depth
$\lambda_{\rm ac}$ in a $6 (w)\times4 (t)\,$mm$^2$ PbIn${10.5\%w}$ slab
in oblique field. It shows the cross-over between the quasistatic
regime ($\Omega/2\pi\lesssim 1$\thinspace kHz) and the free flux-flow
regime ($\Omega/2\pi\gtrsim 1$\thinspace GHz). The 2\thinspace
kHz--20\thinspace MHz experimental range has been achieved by using
two different set-ups, each covering two frequency decades.  Solid
lines are theoretical fits to Eq.~(\ref{admittance})
with $L_{\rm S}=42\pm0.5\,\mu$m and $\rho_{\rm f}=4.0\pm
0.1\,\mu\Omega$cm.  For comparison the Campbell spectrum
(\ref{Campbell}) (dashed-line), with the same set of
parameters, yields a much narrower transition with a maximum
$\lambda^{\prime\prime}/L_{\rm S}=0.272$ larger than the maximum
$\lambda^{\prime\prime}/L_{\rm S}=0.207$ of the solid line.}
\label{depinningspectrum}\end{center}\end{figure}

The experimental procedure is the following: the flux
$\Phi^*\!=\!\Phi(B_0)\!-\!\Phi(0)$ is measured as a function of $B_0$
at constant $\Omega$ up to the normal state where the calibration is
achieved. Then the data for each vortex state are collected as a
function of $\Omega$ to yield the complex spectrum.  In PbIn${10.5\%
wt}$ slabs the resolution was not affected by the reproducibility of
the vortex states. Otherwise we are careful to reproduce the same
vortex state, either by using the same field history or by removing
metastability by applying a large transient over-critical dc current.
The frequency-range is limited to two decades for a given arrangement
of exciting and pick-up coils; it is extended to four decades in
Figs. \ref{depinningspectrum} and \ref{PotPourritArgand}a by
juxtaposing data taken from two distinct experimental setups with 
different windings, with several days lying between the two
measurements. 
 
\section{Skin effect in the mixed state}\label{skin_effect}

Let us first recall the theoretical steps leading to the two-mode
impedance spectrum (\ref{admittance}), developed in
Refs.~\cite{Sonin92,Placais96,Entrup97}.

Consider a superconducting half-space $z<0$, immersed in a constant
magnetic field $B_0\hat z$, and submitted to a small transverse ripple
$b_0\hat x e^{-i\Omega t}$.  In the absence of Hall effects
($\beta^\prime=0$), vortex modes are linearly polarized, and the ac
fields $b_x, E_y, \nu_x, u=\nu_x/{\rm i}k$, \ldots split into two
modes,
\begin{eqnarray} 
b_x &=& \;b_0(\tilde b_1 {\rm e}^{z/\lambda_{\rm f}} + \tilde b_2{\rm
e}^{z/\lambda_{\rm V}}) \; {\rm e}^{-{\rm i}\Omega t},
\label{penetrationB}\\ 
E_y &=& -b_0(\lambda_{\rm f}\tilde b_1 {\rm e}^{z/\lambda_{\rm f}} +
\lambda_{\rm V}\tilde b_2 {\rm e}^{z/\lambda_{\rm V}}) \; i\Omega{\rm
e}^{-{\rm i}\Omega t},
\label{penetrationE}\\ 
\nu_x &=& \;(\nu_1 {\rm e}^{z/\lambda_{\rm f}} +\nu_2 {\rm
e}^{z/\lambda_{\rm V}})\;{\rm e}^{-{\rm i}\Omega t},
\label{penetrationNu}\\ 
u &=& \;(\lambda_{\rm f}\nu_1 {\rm e}^{z/\lambda_{\rm f}} +
\lambda_{\rm V}\nu_2 {\rm e}^{z/\lambda_{\rm V}})\;{\rm e}^{-{\rm
i}\Omega t}. \label{penetrationU}
\end{eqnarray}
with respective wave numbers \cite{Placais96}
\begin{equation}
1/{\rm i}k_1=\lambda_{\rm f}=(1+{\rm i})\delta_{\rm
f}/2,\;\qquad1/{\rm i}k_2=\lambda_{\rm
V}=\lambda\sqrt{\mu_0\varepsilon/(\omega+\mu_0\varepsilon)}.
\label{wave_numbers}\end{equation}

$\tilde b_1$ (resp. $\tilde b_2$) and $\nu_1$ (resp. $\nu_2$) are the
complex reduced amplitudes of the FF-mode (resp. VS-mode).  Field
continuity implies that $\tilde b_1\!+\!\tilde b_2=1$.  Additional
relations between magnetic and vortex fields are obtained by
linearizing the London equation~(\ref{LondonDC}) \cite{Placais96}:
\begin{equation}
\nu_1 = \tilde b_1 \;\frac{b_0}{B_0};\qquad \nu_2 = -\tilde
b_2\;\frac{\omega}{\mu_0\varepsilon}\;\frac{b_0}{B_0}.
\label{amplitudeNu}\end{equation}
Note in Eqs.~(\ref{amplitudeNu}) the different behaviour of the
vortices in the two modes: whereas in the FF-mode vortices follow the
magnetic field ($\omega_1=B_0\nu_1=b_1$), in the VS-mode vortices
incurve in the opposite direction with a deflection drastically
enhanced by the large factor $\omega/\mu_0\varepsilon\simeq
B_0/M\gg1$, where $M$ is the reversible magnetization of the
superconductor.
 
The apparent penetration depth of some combination of two modes is a
weighted average of the individual penetration depths $\lambda_{\rm
f}$ and $\lambda_{\rm V}$:
\begin{equation} 
\lambda_{\rm ac}=\frac{E(0)}{-i\Omega b_0} = \tilde b_1 \lambda_{\rm
f} + \tilde b_2 \lambda_{\rm V}\simeq\tilde b_1\lambda_{\rm f}.
\label{impedance}\end{equation}
Since $\lambda_{\rm V}\lesssim\lambda_{\rm L}=\lambda(\omega=0)$, the
VS-mode can be regarded as superficial on the scale of the sample, so
that its contribution to $\lambda_{\rm ac}$ in (\ref{impedance}) can
be neglected within our experimental resolution ($\sim0.1 \,\mu$m in
$\lambda^\prime$ or $\lambda^{\prime\prime}$).  The surface mode
intervenes not only as a discontinuity in the field amplitude, but
also, and perhaps less intuitively, as a phase-lag ($\arg\tilde b_1$)
in the bulk mode.  The latter effect turns out to be important in
understanding the low-frequency ac penetration.  It will be seen below
that both amplitude and phase of $\tilde b_1$ are controlled by the
boundary conditions for vortices at the surface.

Let us now introduce the boundary conditions for a rough surface and
small vortex displacements ($u\lesssim 1$ \AA) \cite{Entrup97}:
\begin{equation}
\nu_x(0)=\nu_1+\nu_2=- u_1/l.
\label{boundaryconditions}\end{equation}
It should be noted that Eq.~(\ref{boundaryconditions}) is independent
of the frequency.  This comes from the fact that any distortion of the
vortex lattice matching the surface roughness in the small depth
$\lambda_{\rm V}$ can be regarded as quasistatic motion as far as
$\delta_{\rm f}\gg\lambda_{\rm V}$ \emph{i.e.}  in the whole
investigated range of frequencies.  The surface roughness is
characterized by the length $l$.  A geometrical interpretation of $l$
can again be borrowed from superfluids: an isolated vortex, pinned at
the bottom of a spherical bump, when acted on in the bulk (e.g. by a
second-sound wave), bends near the wall so as to keep on ending normal
to the surface; whence a linear relation $\nu_x=-u/l$ with $l=R$, the
radius of curvature of the bump.  Eq.~(\ref{boundaryconditions}) is
nothing but a generalization of this picture to the collective motion
of vortices interacting with their images in the disordered surface.

By extrapolating Eq.~(\ref{boundaryconditions}) for large
displacements $u\sim a$ we obtain the maximum standard deviation of
vortices at the surface, $\bar\nu_x\sim a/l$.  This value of
$\bar\nu_x$ also determines the surface critical-current density
$K_{\rm C}=\varepsilon\bar\nu_x\sim \varepsilon a/l$ according to the
Mathieu-Simon model of the critical state \cite{Mathieu88}.  From
experimental values of the critical current we infer that
$\bar\nu_x\sim 10^{-2}$--$10^{-1}$, and $l\sim$ 1--10$\,\mu$m.

\begin{figure}[hb]
\begin{center}\leavevmode
\includegraphics[width=1\linewidth]{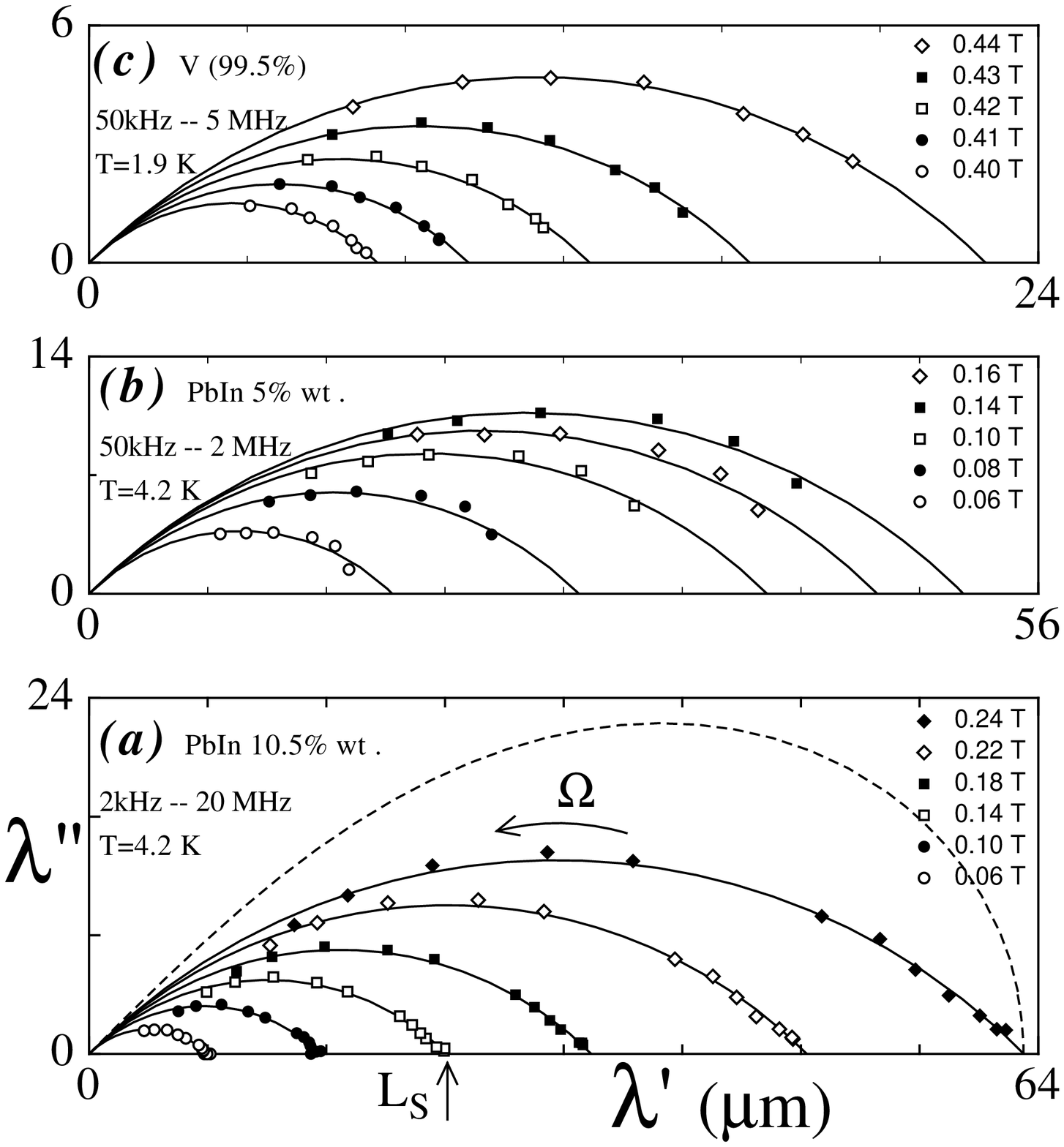}
\bigskip
\caption{The apparent penetration depth $\lambda_{\rm ac}(\Omega)$
displayed in the complex plane.  Data have been measured in the mixed
state of conventional isotropic superconductors: Arc of circles (solid
lines) are fitted to the data according to
Eq.~(\ref{admittance}) by adjusting the skin length $L_{\rm
S}$ (real-axis intercept indicated by an arrow in \emph{(a)}).  A
typical bulk pinning spectrum, given by the Campbell expression
(\ref{Campbell}) with $\lambda_{\rm C}=63\,\mu$m, has been
plotted (dotted line) for comparison to the 0.24T
data. \textbf{\emph{(a)}} a $6(w)\times4(t)\,$mm$^2$ slab of
PbIn($10.5\% wt$) ($\rho_{\rm n}=10.3 \,\mu\Omega$.cm);
\textbf{\emph{(b)}} a $5.8(w)\times2(t)\,$mm$^2$ slab of PbIn($5\%
wt$) ($\rho_{\rm n}=6.5 \,\mu\Omega$.cm); \textbf{\emph{(c)}} a
$6.7(w)\times1(t)\,$mm$^2$ vanadium slab ($\rho_{\rm n}=0.6
\,\mu\Omega$.cm), spark-cut from a cold-rolled commercial plate.}
\label{PotPourritArgand}\end{center}\end{figure}

Using condition (\ref{boundaryconditions}) together with equations
(\ref{penetrationNu}), (\ref{penetrationU}) and (\ref{amplitudeNu}),
we determine $\tilde b_1$ in (\ref{impedance}), which gives the
following expression for the effective penetration depth:
\begin{equation} 
\frac{1}{\lambda_{\rm ac}} = \frac{1}{L_{\rm S}} + 
\frac{1-{\rm i}}{\delta_{\rm f}}, \qquad\mbox{where}\qquad 
L_{\rm S}=l\frac{\omega}{\mu_0\varepsilon}
\label{admittance}\end{equation}
is a length directly related to the roughness length $l$, and
therefore connected with the order of magnitude of critical currents
by $K_{\rm C}\sim \omega a/\mu_0L_{\rm S}$.

The frequency dependence of $\lambda_{\rm ac}$ is governed by that of
the flux-flow skin depth $\delta_{\rm f}\propto 1/\sqrt{\Omega}$.  The
right-hand side of Eq.~(\ref{admittance}) is a straight line in the
complex plane, which, by inversion, gives a quarter-circle for the
spectrum $\lambda_{\rm ac}(\Omega)$ (Fig.~\ref{PotPourritArgand}).
Through the action of the boundary conditions, it turns out that, in
the low-frequency limit, $\lambda_{\rm ac}\simeq\tilde b_1\lambda_{\rm
f}\simeq L_{\rm S}$ is real and independent of $\Omega$.  The length
$L_{\rm S}$, as $\lambda^\prime(0)$, plays the role of the Campbell
length.  We stress however that, in contrast to $\lambda_{\rm C}$,
$L_{\rm S}$ \emph{is not an actual penetration depth}, but rather some
fraction of a much longer depth $\delta_{\rm f}$.

\begin{figure}[ht]
\begin{center}\leavevmode
\includegraphics[width=0.8\linewidth]{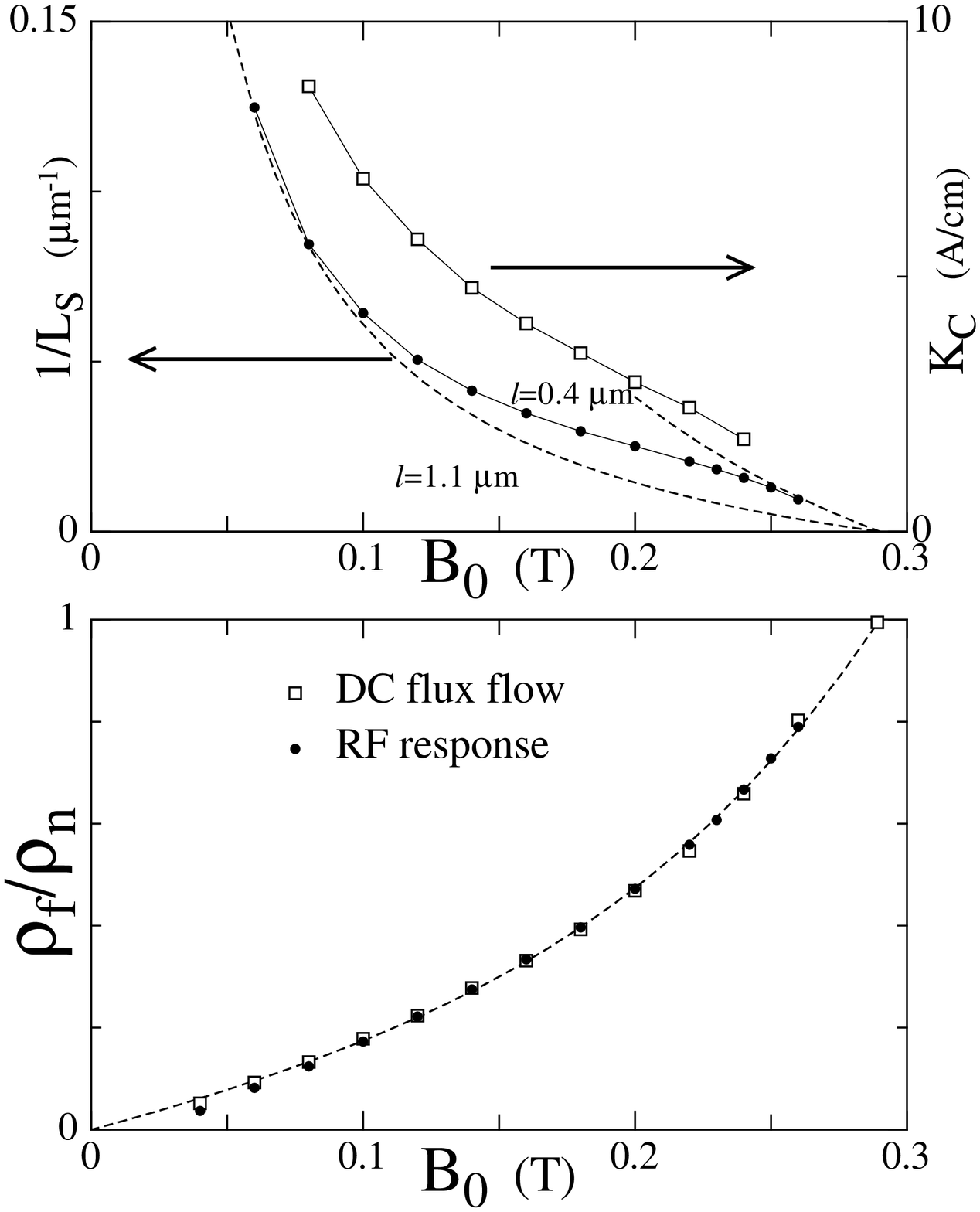}
\bigskip
\caption{\emph{Upper panel} : Typical field-dependences for the
skin-length $L_{\rm S}$ (bullets) and the surface critical current
density $K_{\rm C}$ (squares) in PbIn$_{10.5\%w}$ ($B_{\rm
c2}=0.29\,$T, $\kappa\simeq3.5$) at $T=4.2\,$K.  Dotted lines are
theoretical curves $L_{\rm S}=l B_0/\mu_0\varepsilon$ using the
Abrikosov expression $\varepsilon\simeq(B_{\rm
c2}-B_0)/2.32\mu_0\kappa^2$, and a pinning length $l=1.1 \,\mu$m
(resp.  $l=0.4 \,\mu$m) for intermediate fields (resp. high fields).
The product $K_{\rm C} L_{\rm S}\sim 10\,$mA agrees well with the
theoretical estimate $\sqrt{B\varphi_0}/\mu_0$ ($= 14$ mA at
$B_0=0.15$ T) for surface pinning.  \emph{Lower panel} : the field
dependence of the flux flow resistivity in PbIn${10.5\%w}$ at $T=4.2\,$K
deduced from the slope of dc voltage-current characteristics
(squares) and from rf surface impedance (bullets). rf data are deduced
from the frequency dependence of $\lambda_{\rm ac}(\Omega)$ in
Fig.~\ref{PotPourritArgand}a.}
\label{LSandRhoFdeB}\end{center}\end{figure}

Fig.~\ref{depinningspectrum} shows a typical complex spectrum of
$\lambda_{\rm ac}$ in a $\log$-frequency scale. Compared to our
previous measurements \cite{Entrup97}, the number of frequency points
and the frequency span have been doubled so as to cover most of the
depinning transition.  We confirm our observation that the cross-over
is much broader than predicted by the Campbell bulk-pinning expression
(\ref{Campbell}) (dotted lines in the figure) and extends over 6
decades of frequency ($10^3$--$10^9$ Hz in
Fig.\ref{depinningspectrum}).  This observation is in apparent
contradiction with that of Gittleman and Rosenblum \cite {Gittleman66}
who reported a depinning transition in the ac resistivity of similar
materials, over 2-decades of frequency. In fact, there is no
inconsistency if one takes into account the finite thickness of their
samples (see Sec.~\ref{size_effects} below).  The quality of the fit
to Eq.~(\ref{admittance}) (solid line) in Fig.\ref{depinningspectrum}
is, in itself, a strong indication for the two-mode electrodynamics.
Apart from theoretical considerations, formula (\ref{admittance})
proves powerful in extracting reliable values of the free-flow
resistivity $\rho_{\rm f}$ within a few percent.  This method compares
favorably with microwave techniques.

The universal shape of the depinning spectra is well illustrated in
the Argand diagram of Fig.~\ref{PotPourritArgand}, where data from two
PbIn alloys ($\rho_{\rm n}=6.5$ and $10.3 \,\mu\Omega$cm
respectively)) and a vanadium sample ($\rho_{\rm n}=0.6
\,\mu\Omega$cm) have been collected.  The characteristic
quarter-circle is observed in a variety of experimental conditions and
samples, including very pure niobium ($\rho_{\rm n}\lesssim
10\,$n$\Omega$cm) in Ref.~\cite{Entrup97}.  Two distinctive features,
well observed in Fig.~\ref{PotPourritArgand}\emph{a}, are worth
mentioning : {\bf i)} the linear behavior
$\lambda^{\prime}\!+\!\lambda^{\prime\prime}\!=\!\lambda^\prime(0)$ at
low frequency, and {\bf ii)} the maximum of $\lambda^{\prime\prime}$,
$\lambda^{\prime\prime}=\frac{1}{2}\tan(\pi/8)\lambda^\prime(0)=
0.207\lambda^\prime(0)$, which is $30\%$ less than the generally
accepted classical value $0.272\lambda^\prime(0)$.

The field dependence of both $L_{\rm S}$ and critical current
densities $K_{\rm C}$ are shown in Fig.~\ref{LSandRhoFdeB} (upper).
The order of magnitude of the roughness length $l$ can be obtained by
using the Abrikosov expression $\mu_0\varepsilon=(B_{\rm
C2}\!-\!\omega)/2.32\kappa^2$ in $1/L_{\rm S} =
\mu_0\varepsilon/l\omega$ and $\omega\!=\!B_0$.  With $\kappa=3.5$ we
find reasonable values $0.4<l<1\,\mu$m (dashed lines in
Fig.~\ref{LSandRhoFdeB}).  Moreover, $K_{\rm C}=I_{\rm C}/2w$ values,
deduced from dc measurements, are also found to agree with the
estimate $K_{\rm C}\approx \omega a/\mu_0L_{\rm
S}$. Fig.~\ref{LSandRhoFdeB} (lower) illustrates the resolution in
$\rho_{\rm f}(\omega)$ achieved with rf surface impedance.
 
\begin{figure}[ht]
\begin{center}\leavevmode
\includegraphics[width=0.9\linewidth]{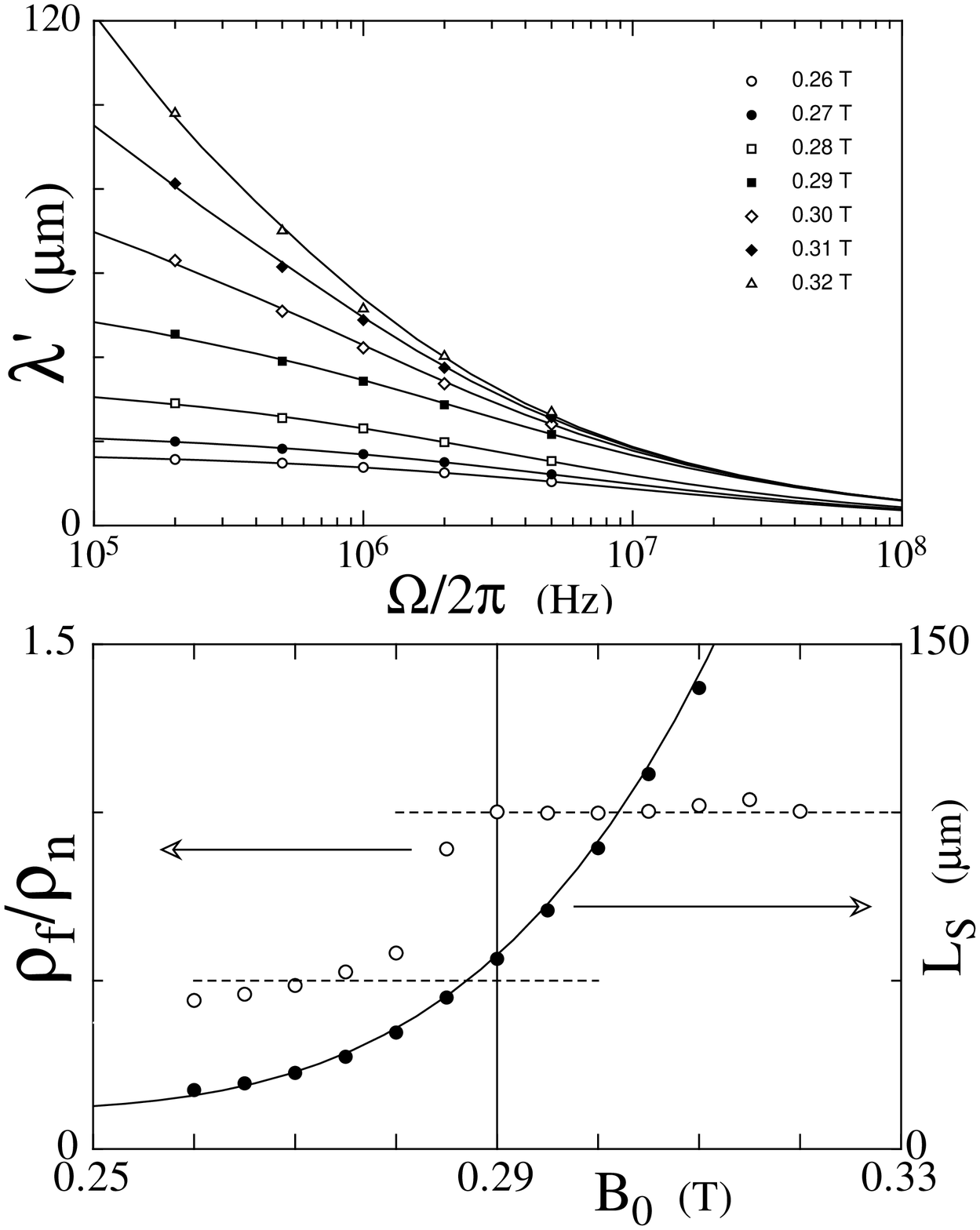}
\bigskip
\caption{\emph{Upper panel}: Continuity of the spectra $\lambda_{\rm
ac}(\Omega)$ across the bulk-superconductivity transition in oblique
field.  Data are taken from a $6.1 (w)\times 1.22 (t)\,$mm$^2$
PbIn${10.5\% wt}$ at $T=4.2\,$K ($B_{\rm C2}=0.29\,$T).  In this geometry,
currents flow at 45 degrees to the magnetic field, and the penetration
of the ac-field is symmetric over the four faces of the slab
throughout the investigated range $0.9\leq B_0/B_{\rm C2}\leq 1.1$.
Agreement with the theoretical spectrum (\ref{admittance}) is observed
on both sides of the transition which is a strong indication that the
screening mechanism is the same in both cases.  \emph{Lower panel} :
The skin-length $L_{\rm S}(B_0)$ ($\bullet$) and the flux-flow
resistivity $\rho_{\rm f}(B_0)$ ($\circ$), deduced from the fits of
$\lambda_{\rm ac}$.  $L_{\rm S}(B_0)$ is itself continuous through the
transition, as is the dc critical current.  The only accident observed
at $B_{\rm C2}$ is a drop by a factor 2 in $\rho_{\rm f}(\pi/4)$ which
signals the vanishing of the mixed-state anisotropy $\rho_{\rm
f}(\theta)=\rho_{\rm f}(0)\cos^2\theta$.}
\label{suprasurface}\end{center}\end{figure}

\section{Skin effect through the superconducting sheath}
\label{sheath_superconductivity}

We have investigated the field range $B_0/B_{\rm C2}=$ 0.9--1.1 to
explore the behaviour of the ac response in the transition from the
bulk mixed state to the surface superconductivity.  As surface
superconductivity is not destroyed in oblique field, we took advantage
of this geometry to symmetrize the role of the four faces of the
sample (see Sec.~\ref{set-up}).  We recall that the superconducting
sheath (depth $\sim\xi$) is populated by short vortices (flux spots)
with a density $\omega=B_{0\perp}$ (see Ref.~\cite{Mathieu93} and
references therein).

The most striking fact in Fig.~\ref {suprasurface} is that, at the
transition, there is no change in the shape of spectra, which, both
above and below $B_{\rm C2}$, obey the theoretical expression
(\ref{admittance}).  Above $B_{\rm C2}$ the bulk is normal, and the
existence of screening surface currents is obvious; therefore, the
two-mode electrodynamics and the free bulk penetration of the wave,
which underlie Eq.~(\ref{admittance}), are not surprising.
 
Concerning the field-dependent parameters involved in
Eq.~(\ref{admittance}), $L_{\rm S} $ and $\rho_{\rm f}$, we observe no
discontinuity of $L_{\rm S}(B_0)$, whereas a jump is observed in
$\rho_{\rm f}$ as shown in Fig.~\ref {suprasurface}b. The
continuity of $L_{\rm S}(B_0)$ is consistent with the observation that
critical currents decrease monotonically through the transition
together with the strength of the surface vortex state.  The vanishing
of the vortex structure in the bulk at $B_{\rm C2}$ results in an
abrupt jump from $\rho_{\rm n}\cos^2\theta=\rho_{\rm n}/2$ to
$\rho_{\rm n}$ as expected from the anisotropy of resistivity in the
mixed state \cite{Mathieu93}.

\section{Size effects in thin slabs}\label{size_effects}

Equation~(\ref{admittance}), and results in the above Section,
concerned the ac response of two independent half spaces.  In thin
slabs and/or at low frequencies, fields penetrating from opposite
faces interfere in the bulk whenever $t\lesssim 2\delta_{\rm f}$.
Assuming symmetric surface conditions on both faces $z=\pm t/2$, the
corresponding one-dimensional solution takes the even form :

\begin{eqnarray} 
\frac{b}{b_0} = \;\tilde b_1 \frac{\cosh(z/\lambda_{\rm f})} 
{\cosh(t/2\lambda_{\rm f})} &+& \;\tilde
b_2\frac{\cosh(z/\lambda_{\rm V})}{\cosh(t/2\lambda_{\rm V})}
\label{platepenetrationB}\\
\frac{E}{-i\Omega b_0} = \tilde b_1 \lambda_{\rm f}
\frac{\sinh(z/\lambda_{\rm f})}{\cosh(t/2\lambda_{\rm f})} &+&
\tilde b_2 \lambda_{\rm V}\frac{\sinh(z/\lambda_{\rm V})}
{\cosh(t/2\lambda_{\rm V})}
\label{platepenetrationE}\\
\frac{B_0\nu}{b_0} = \;\tilde b_1 \frac{\cosh(z/\lambda_{\rm f})}
{\cosh(t/2\lambda_{\rm f})} &-& \frac{\omega}{\mu_0\varepsilon}
\tilde b_2\frac{\cosh(z/\lambda_{\rm V})}{\cosh(t/2\lambda_{\rm V})}
\label{platepenetrationNu}\\
\frac{B_0u}{b_0} = \tilde b_1\lambda_{\rm f}
\frac{\sinh(z/\lambda_{\rm f})}{\cosh(t/2\lambda_{\rm f})}& -&
\frac{\omega}{\mu_0\varepsilon} \tilde b_2\lambda_{\rm V}
\frac{\sinh(z/\lambda_{\rm V})}{\cosh(t/2\lambda_{\rm V})}
\label{platepenetrationU}\end{eqnarray}

The effective penetration depth reads
\begin{equation} 
\lambda_{\rm ac}=\frac{E(t/2)-E(-t/2)}{-2{\rm i}\Omega b_0} = 
\tilde b_1 \lambda_{\rm f}\;{\tanh(t/2\lambda_{\rm f})} + 
\tilde b_2\lambda_{\rm V} \simeq \tilde b_1 \lambda_{\rm f} 
{\tanh(t/2\lambda_{\rm f})}.
\label{plateimpedance}\end{equation}
Thin films ($t\lesssim 2\lambda_{\rm V}$) are excluded here by
assuming ${\tanh(t/2\lambda_{\rm V})}=1$.  Inserting boundary
conditions $\nu(\pm t/2)=\mp u_1(\pm t/2)/l$, Eq.~(\ref{admittance})
becomes
\begin{eqnarray} 
\frac{1}{\lambda_{\rm ac}}&=&\frac{1}{L_{\rm S}} +
\frac{1-i}{\delta_{\rm f}}\coth{\frac{(1-i)t}{2\delta_{\rm f}}}
\label{plateadmittance}\\ \frac{1}{\lambda_{\rm ac}}&\simeq
&\frac{1}{L_{\rm S}} + \frac{2}{t} \qquad\qquad(t/\delta_{\rm f}
\rightarrow 0).
\label{thinplateadmittance}\end{eqnarray}

\begin{figure}[ht]
\begin{center}\leavevmode
\includegraphics[width=1\linewidth]{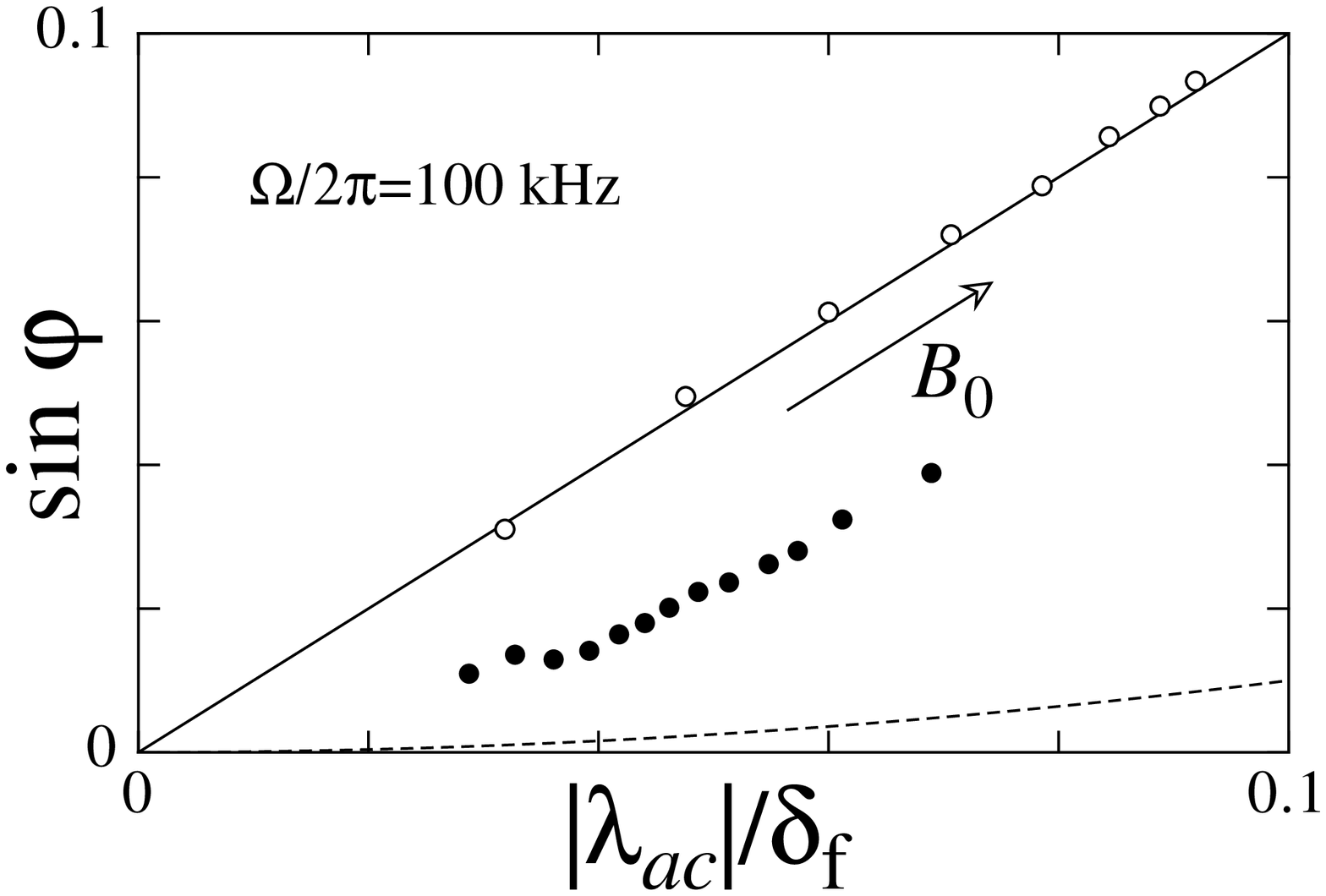}
\bigskip
\caption{The sine of the loss-angle
$\sin\varphi=\lambda^{\prime\prime}/|\lambda_{\rm ac}|$ as a function
of $|\lambda_{\rm ac}|/\delta_{\rm f}$ in two slabs of the same
PbIn${10.5\%wt}$ material with different thickness $t=0.3\,$mm
($\bullet$) and $t=1.3\,$mm ($\circ$).  The skin-depth $\delta_{\rm
f}$, or the flux-flow resistivity $\rho_{\rm f}$, is deduced from
separate dc voltage-current characteristics.  Data points
($|\lambda_{\rm ac}|,\varphi$) have been taken at a constant low
frequency as a function of magnetic field : $B_0=0.04$--$0.18\,$T and
$\delta_{\rm f}=0.13$--$0.32\,$mm in the $t=1.3\,$mm slab;
$B_0=0.06$--$0.27\,$T and $\delta_{\rm f}=0.15$--$0.45\,$mm in the
$t=0.3\,$mm slab.  The theoretical law $\sin\varphi = |\lambda_{\rm
ac}|/\delta_{\rm f}$ for the half space is well obeyed in the thick
slab with $\delta_{\rm f}\lesssim t/4$, while a significant deficit in
the losses of thin slab is observed for $\delta_{\rm f}\simeq
0.66$--$2 t$.  Note that, $|\lambda_{\rm ac}|\simeq 5$--$30 \,\mu$m
and $t=10$--$60 |\lambda_{\rm ac}|$ in the same field range, so that
no size effect would be expected if $\lambda_{\rm ac}$ was the actual
penetration depth.  Quantitative comparison with theory is difficult
in this thin limit as losses become drastically sensitive to the
symmetry of surface conditions (compare Eq.~(\ref{plateimpedance}) with
Eq.~(\ref{odd_admittance})). }
\label{thinslablossess}\end{center}\end{figure}

Comparing Eq.~(\ref{plateadmittance}) with Eq.~(\ref{admittance}), it
is seen that size effects will start when $t$ is decreased below about
$ 3\delta_{\rm f}$, where the $\coth$ factor deviates significantly from
unity. As a result losses are significantly reduced, the loss angle
$\varphi$ falling off below the half-space level given by
$\sin\varphi\!=\!|\lambda_{\rm ac}|/\delta_{\rm f}$.  Data ($\varphi,
|\lambda_{\rm ac}|$) obtained at 100\thinspace kHz in PbIn slabs,
where $\rho_{\rm f}(B_0)$ (then $\delta_{\rm f}$) had been separately
measured, clearly demonstrate these size effects (see
Fig.\ref{thinslablossess}).  The fact that for decreasing sample
thicknesses size effects are encountered when $t\sim\delta_{\rm f}$,
\emph{i.e.} well before $t\sim L_{\rm S}$, is not consistent with any
one-mode theory (see Sec.~\ref{Introduction}).

According to Eq.~(\ref{thinplateadmittance}), a second size effect
occurs in the thin limit when $t\sim 2L_{\rm S}$. This second
``dimensional'' cross-over, which affects now the real part
$\lambda^\prime$, could wrongly suggest that $L_{\rm S}$ is a true
length in accordance with one-mode approaches. This shows that one
must be careful in interpreting susceptibility measurements.

Let us conclude by briefly examining resistivity measurements in foils
within the thin limit.  We refer in this respect to the pioneering
work of Gittleman and Rosenblum \cite{Gittleman66}, as also to recent
measurements in YBaCuO platelets \cite{Wu97}.  The analysis of such
resistivity spectra is much more intricate.  Many authors, using this
method assume that the current density does not vary along the width
$w$ of the foil; this requires the absence of two-dimensional
skin-effects, $\delta_{\rm f}\gg w>t$, a condition that is hardly
fulfilled over several frequency decades.

The one-dimensional treatment of the ac resistivity relies on the odd
solution of Eqs.~(\ref{penetrationB})--(\ref{penetrationU}). The
effective resistivity of the foil is defined as
\begin{equation}
\rho_{\rm ac}=\mu_0 \frac{E(t/2)}{b_0}\;\frac{t}{2} =-{\rm
i}\mu_0\Omega\lambda_{\rm ac}\; \frac{t}{2},
\label{resistivity}\end{equation} 
where
\begin{equation} 
\frac{1}{\lambda_{\rm ac}}= \frac{1}{L_{\rm S}} +
\frac{1-i}{\delta_{\rm f}}\tanh{\frac{(1-i)t}{2\delta_{\rm f}}}.
\label{odd_admittance}\end{equation}

In the limit $t\ll2\delta_{\rm f}$,
Eqs.~(\ref{resistivity}),(\ref{odd_admittance}) reduce to :
\begin{equation} 
\frac{-{\rm i}\mu_0\Omega}{\rho_{\rm ac}} = \frac{2}{t\lambda_{\rm ac}}=
\frac{2}{tL_{\rm S}} -  \frac{2i}{\delta_{\rm f}^2}.
\label{thin_resistivity}\end{equation}
Surprisingly enough, we find the same resistivity spectrum as that
given by Eq.~(\ref{Campbell}) which follows from the bulk-pinning
theory.  Both Eqs.~(\ref{Campbell}) and (\ref{thin_resistivity})
equally fit the Gittleman-Rosenblum data (by setting $\lambda_{\rm
C}=\sqrt{L_{\rm S}t/2}$ in Eq.~(\ref{thin_resistivity})).  We are led
to the conclusion that resistivity experiments alone cannot decide
between very contrasted models of pinning.

\section{Conclusions}\label{conclusion}

We have reported the first comprehensive investigation of dynamic
depinning transition in the surface impedance geometry.  It relies on
the description of the depinning transition through
Eq.~(\ref{admittance}) and requires the use of thick samples
($t\gtrsim 3 \delta_{\rm f}$).  The agreement is so good that
surface-impedance measurements in the rf-domain provide a precise and
accurate way of measuring both parameters $L_{\rm S}$ and $\rho_{\rm
f}$.  As shown in Fig.~\ref{LSandRhoFdeB}, dc data on the flux-flow
resistivity, when available, confirm the accuracy of the method; in
other cases where low-level ac techniques are necessary (large
critical currents, ultra-low temperatures, \ldots), the surface
impedance becomes the best suited non-invasive method for measuring
$\rho_{\rm f}$.  Apart from this metrological aspect, it is clear
that, as concerns the pinning length $L_{\rm S}=\lambda^\prime(0)$,
our interpretation is completely at variance with classical views on
pinning (\cite{Tinkham96} p. 345).  In this respect we wish to add
some comments.

The remarkable agreement with the two-mode spectrum (\ref{admittance})
as well as the observation of predicted size effects for
$t\sim\delta_{\rm f}$ at low frequencies bring out two important
facts: {\bf i)} The effective length, as deduced from the surface
inductance $X/\Omega \!=\!\mu_0\lambda^\prime(0)$, is \emph{not} an
actual penetration depth; {\bf ii)} free flow is observed in the bulk
(over thicknesses $t\gtrsim 1$ mm), in spite of a large density of
volume defects in a variety of polycrystalline samples from alloys to
pure metals.

In the literature, $\lambda^\prime(0)$, usually referred to as the
Campbell penetration depth, is directly expressed in terms of the
bulk-pinning Labusch parameter $\alpha_{\rm L}$ (see
Sec.~\ref{Introduction}).  In soft samples, such as used in this work,
bulk pinning is commonly analysed in the frame of the collective
pinning theory of Larkin-Ovchinnikov \cite{Tinkham96}, where
$\alpha_{\rm L}$ is in turn related to Larkin transverse and
longitudinal correlation lengths $R_{\rm C}$ and $L_{\rm C}$. These
correlations lengths are intricate expressions of thermodynamic and
pinning parameters (elastic constants $c_{44}$ and $c_{66}$, core
radius $\xi$, pinning-centers concentration $n$ and the individual
pinning force $f$).  For thick samples (3D collective pinning) one
expects that $\lambda^\prime(0)\approx L_{\rm C}\sim
c_{44}c_{66}\xi^2/nf^2$ \cite{Beek93}. Thus many authors consider the
lengths $\lambda^\prime(0)$ deduced from the low-frequency
susceptibility as a routine measurement of $L_{\rm C}$ and merely
investigate the field and temperature dependence of $L_{\rm C}$.  This
studies must be taken with due caution; our results show that a
spectral analysis should precede any interpretation.

The role of point defects and elastic constants is certainly important
in analyzing the short and long-range distortions of the vortex
lattice, as observed in a number of techniques (Bitter decoration,
neutron scattering, STM, \ldots).  However, our experiments seriously
question the relevance of this mechanism of weak pinning in transport
properties of soft samples.  Here, we indeed exclude the case of hard
materials where additional interfaces in the bulk may, to some extent,
act like the surfaces of soft samples (twin planes in YBCO
\cite{Maggio_Aprile97}, precipitates in industrial wires, \ldots).

The existence of a free vortex flow in the bulk, and the exclusive
role of the surface in the critical-current properties of soft
superconductors, have been confirmed in previous experiments under
over-critical conditions: for instance the voltage and field noise in
the flux-flow regime \cite{Placais94}, the high-level step-response of
a slab in parallel field \cite{Vasseur98}, or surface localization of
sub-critical currents \cite{Thorel73}, \ldots We emphasize that these
experiments have been specially designed to discriminate surface and
bulk currents. It is clear that the existence of bulk sub-critical
currents would imply bulk pinning, but there is up to now no
experimental evidence for such bulk currents in soft samples. In
particular one cannot, in this respect, rely on the recent Hall probe
technique, which determines the radial distribution of the current in
platelets. The observed pile-of-sand profile of the magnetic field
confirms the critical state model \cite{Abulafia95}, but critical
state model should not be confused with bulk-pinning model.  The
critical-current densities in this geometry may be superficial and the
technique used cannot resolve the axial distribution.

\end{document}